\documentclass[prd,showpacs,floatfix,twocolumn,,amsmath,amssymb,floatfix]{revtex4}
\usepackage{graphicx,color,dcolumn,booktabs,bm}
\usepackage{longtable,lscape}
\usepackage{txfonts}
\usepackage{overpic}
\usepackage{amssymb}
\usepackage{indentfirst}
\usepackage{epsfig} 
\usepackage{epstopdf}   
\usepackage{slashed}  
\usepackage{cases}
\usepackage{color}
\usepackage{multirow}
\usepackage{hhline}
\usepackage[section]{placeins}
\usepackage{graphicx,color,dcolumn,booktabs,bm}

\graphicspath{{Figures/}} %

\usepackage[colorlinks, citecolor=blue,anchorcolor=red,menucolor=red, linkcolor=red,filecolor=red,runcolor=red,urlcolor=blue,frenchlinks=red]{hyperref}

\usepackage[none]{hyphenat} 

\usepackage{color,xcolor,fancybox,epsf,rotating,colordvi}

\newcommand{\tpz}{{^3P_0}}
\newcommand{\kstarz}{{K_0^*}}

\newcommand{\pbs}{{\boldsymbol{p}}}

\begin{document}
	
	\title{Canonical interpretation of the newly observed $J^P =1^+$ structure $X(2085)$}
	
	\author{Tian-Ge Li}
	\affiliation{Joint Research Center for Theoretical Physics, School of Physics and Electronics, Henan University, Kaifeng 475004, China}
	
	\author{Sheng-Chao Zhang}
	\affiliation{Joint Research Center for Theoretical Physics, School of Physics and Electronics, Henan University, Kaifeng 475004, China}

	\author{Guan-Ying Wang}\email{wangguanying@henu.edu.cn}
	\affiliation{Joint Research Center for Theoretical Physics, School of Physics and Electronics, Henan University, Kaifeng 475004, China}

\author{Qi-Fang L\"u}
\email{lvqifang@hunnu.edu.cn} %
\affiliation{Department of Physics, Hunan Normal University, and Key Laboratory of Low-Dimensional Quantum Structures and Quantum Control of Ministry of Education, Changsha 410081, China}
\affiliation{Key Laboratory for Matter Microstructure and Function of Hunan Province, Hunan Normal University, Changsha 410081, China}

	\date{\today}
	
	\begin{abstract}
		Inspired by the newly observed $X(2085)$ by the BESIII Collaboration, we study the strong decay behaviors of excited axialvector strange mesons within the quark pair creation model. Our results indicate that the $K_1(1793)/K_1(1861)$ can be regarded as the same $K_1(2P)$ state, and the $K_1(1911)$ is assigned as the $K_1(2P^\prime)$ state. Considering the mass, spin-parity, and decay behaviors, we interpret the newly observed $X(2085)$ as the radially excited $K_1(3P)$ state,  which mainly decays into the $\rho(1450) K$, $\omega(1420)K$, $\pi K^*(1410)$, $\rho K_1(1270)$, and $\rho K^*(892)$ final states. Also, the width of $K_1(3P^\prime)$ state is predicted to be about 300 MeV, which can be searched for by future experiments.  We expect that present calculations can help us to better understand the nature of the $X(2085)$ structure.
	\end{abstract}
	\pacs{ }
	\maketitle

	\section{Introduction}{\label{Introduction}}
	
Recently, the  BESIII Collaboration reported the signal of $X(2085)$ in the process $e^+e^-$ $\to$ $pK^{-} \bar{\Lambda}$ + $c.c$ with a statistical significance greater than $20\sigma$. The quantum numbers of this new structure are determined as $J^P =1^+$ in an amplitude analysis, and the pole positions are measured to be $M_{pole} = (2086\pm4 \pm 6)$ MeV and $\Gamma_{pole} = (56 \pm 5 \pm 16)$ MeV~\cite{BESIII:2023xxx}. Since the X(2085) locates near the $p\bar \Lambda$ threshold, it displays as a clear enhancement in the $p\bar \Lambda$ invariant mass distribution. Although the isospin is not determined experimentally, the strong decay mode $p\bar \Lambda$ suggests that it should has an isospin $I=1/2$ in theory. Considering the mass, spin-parity, and decay mode, one can investigate the possible nature of this structure, such as a conventional strange meson, an exotic state, the mixture between conventional and exotic states, or the kinematic effect near threshold.

Understanding the mass spectra and internal structures for various hadrons is the basic topic of hadron physics. Besides the conventional mesons and baryons, more and more exotic resonances are observed in the past two decades, which challenge the traditional $q\bar q$ and $qqq$ pictures in the quark model and provide new insights  for the strong interactions in the non-perturbative energy region. However, before claiming a particle as an exotic state, the possibility of it being an ordinary meson or baryon should be carefully studied. Thus, it is essential for us to  study the conventional spectroscopy carefully and perform the reasonable interpretation for the newly observed particles.

There have been already lots of works for the low-lying strange mesons. Godfrey and Isgur calculated the mass spectrum of strange mesons using the relativized quark model~\cite{Godfrey:1985xj}, which can describe the low-lying spectrum reasonably. The strong decays and mixing angles for $K_1(1270)$ and $K_1(1400)$ are studied within the quark pair creation model together with the relativized quark model~\cite{Blundell:1995au}. Then, Barnes et al. investigated the strong decays of the observed excited kaons by using the quark pair creation model associated with the simple harmonic oscillator wave functions~\cite{Barnes:2002mu}. Ebert et al. analyzed the mass spectrum and Regge trajectories of kaons within a  relativistic quark model~\cite{Faustov:2009dlw}, where many excited kaons are predicted theoretically. Pang et al. investigated the low-lying mass spectrum and strong decays for strange mesons systematically~\cite{Pang:2017dlw}.  In our previous work~\cite{Wang:2022pxm}, we investigated the $0^{++}$ sector for strange mesons with the quark pair creation model, and predicted possible highly excited ones. A recently systematical investigation on excited kaons was done, which provides reasonable information for excited strange mesons~\cite{Oudichhya:2023lva}. However, owing to the complexity of strange mesons, it is difficult to perfectly explain all the experimental data simultaneously. Until now, the low-lying strange mesons are far away form established, much less the highly excited states. 

Here, we concentrate on the axialvector strange mesons, which has a spin-parity $J^P =1^+$. With unequal quark masses, the total spin of two quarks $\boldsymbol{S}=\boldsymbol{s_1}+\boldsymbol{s_2}$ is not a conserved quantity, that is $[\boldsymbol{S}, H] \neq 0$. Then, the physical states should be the superpositions of the theoretical states $n^{2S+1}L_J$, where the notation of spectroscopy is adopted. These superpositions introduce the mixing angles, and increase the difficulties and uncertainties of theoretical studies. For convenience, the observed $J^P =1^+$ strange mesons together with possible assignments are listed in Table~\ref{tab:prefer}. Actually, except for the lowest $K_1(1270)$ and $K_1(1400)$ resonances, the assignments of other observed $J^P =1^+$ strange mesons are still unclear so far.  We want to emphasize that the study on these excited strange mesons can help us to establish and  perfect the excited nonets, understand the spectroscopy of light mesons systematically, supply a clear physical background for the investigation of exotic states, and provide helpful information for future experiments in searching for highly excited particles.  

In present work, we adopt the quark pair creation model to study the strong decay behaviors of $J^P =1^+$ strange mesons, and attempt to clarify the nature of $X(2085)$ and relevant resonances. Our results indicate that the $X(2085)$ can be regarded as the conventional $K_1(3P)$ state in the traditional quark model. Also, the width of $K_1(3P^\prime)$ state is predicted to be about 300 MeV, which can be searched for by future experiments.  

	\begin{table}[h]
		\centering
		\caption{The experimental status and possible assignments for the axialvector kaons.}
		\label{tab:prefer}
			\begin{tabular*}{8.5cm}{cccc}
			\hline\hline
			State        &Mass (MeV)   &Width (MeV)   & Possible assignments \\
			\hline
         $K_1(1270)$   ~\cite{Zyla:2020zbs}   & $1253\pm7$  & $90\pm20$      &  $K_1(1P)$       \\
          $K_1(1400) $~\cite{Zyla:2020zbs}       & 1403$\pm$7  & 174$\pm$13     &$K_1(1P^\prime) $          \\
           $K_1(1650) $~\cite{Zyla:2020zbs}     & 1650$\pm$50  & 150$\pm$50      &  $\cdot \cdot \cdot$                  \\
			$K_1(1861)$~\cite{LHCb:2021xxx}        & 1861$\pm$10$^{+16}_{-46}$  & 149$\pm$41$^{+231}_{-23}$    &$K_1(2P) $  \\
			$K_1(1911)$~ \cite{LHCb:2021xxx}   & 1911$\pm$37$^{+124}_{-48}$  & 276$\pm$50$^{+319}_{-159}$         &$K_1(2P^\prime) $      \\
			$K_1(1793)$~\cite{LHCb:2016xxx}  & 1793$\pm$59$^{+153}_{-101}$  & 365$\pm$157$^{+138}_{-215}$  &$K_1(2P)$    \\
			$K_1(1968)$ ~\cite{LHCb:2016xxx}   & 1968$\pm$65$^{+70}_{-172}$  & 396$\pm$170$^{+174}_{-178}$      & $\cdot \cdot \cdot$        \\
				$X(2085)$~\cite{BESIII:2023xxx}          & 2086$\pm$4$\pm$6 &  56$\pm$5$\pm$16                               & $K_1(3P)$    \\
			\hline\hline
		\end{tabular*}
	\end{table}	

This paper is organized as follows. In Sec.~\ref{sec:Mass}, we analyse the masses of possible candidates for the radially excited $J^P =1^+$ strange mesons. In Sec.~\ref{sec:Decay}, we briefly introduce the quark pair creation model, and then present our results and discussions. Finally, a summary is given in Sec.~\ref{sec:summary}.
 
\section{Mass}{\label{sec:Mass}}

In this section, we investigate the possible assignments of the observed axialvetor strange mesons according to their masses. In the Review of Particle Physics (RPP)~\cite{Zyla:2020zbs}, the $K_1(1270)$ and $K_1(1400)$ are clarified into the mixtures or superpositions of  $K_1(1^1P_1)$ and $K_1(1^3P_1)$ states. More explicitly, the mixing scheme for physical states for $1^+$ strange mesons can be expressed as
 \begin{eqnarray}
	\begin{pmatrix}
		~~|K_1(nP)\rangle~  \\
		~~|K_1(nP^\prime)\rangle~  \\
		
	\end{pmatrix}
	=
	\begin{pmatrix}
		~{\rm cos}~\theta_{nP} & {\rm sin}~ \theta_{nP}~ \\
		~-{\rm sin}~ \theta_{nP} & {\rm cos}~ \theta_{nP}~ \\
	\end{pmatrix}
	\begin{pmatrix}
		~~|K_1(n^1P_1)\rangle~  \\
		~~|K_1(n^3P_1)\rangle~  \\
	\end{pmatrix},
\end{eqnarray}
where the $\theta_{nP}$ is  known as the mixing angle between $K_1(n^1P_1)$ and $K_1(n^3P_1)$ states. These mixing angles will be discussed in the next section together with the strong decays.

The next $1^+$ resonance listed in the RPP is named as $K_1(1650)$~\cite{Zyla:2020zbs}, which collects several experimental measurements and assigns the different structures as a single state $K_1(1650)$. Experimentally, the original $K_1(1650)$ was observed in the $K^+p \to \phi K^+p$ reaction by using the CERN Omega spectrometer in 1986~\cite{Frame:1985ka}.  The $K_1(1793)$ and $K_1(1968)$ were first reported in the $J/$$\psi$$\phi$ invariant mass distribution of the $B^+$ $\to$ $J$/$\psi$$\phi$$K^+$ reaction by LHCb Collaboration in 2016~\cite{LHCb:2016xxx}. In 2021,  the LHCb Collaboration reanalysed the $B^+$ $\to$ $J/$$\psi$$\phi$$K^+$, and  observed the $K_1(1861)$ and $K_1(1911)$ resonances~\cite{LHCb:2021xxx}. Because these structures have quite different masses and decay behaviors, it is difficult to treat them as a single state in theory. Thus, we prefer to denote them as five different structures  and discuss the assignments separately as shown in Table~\ref{tab:prefer}. 

Theoretically, there should have two states $K_1(2P)$ and $K_1(2P^\prime)$ in the mass region of $1750 \sim 1950~\rm{MeV}$. We estimate a wide range for $2P$-wave states since the theoretical works also predicted significantly different masses for them. The Godfrey-Isgur's relativized quark model predicted the $K_1(2P)$ and $K_1(2P^\prime)$ states to be 1897 and 1928 MeV, respectively~\cite{Godfrey:1985xj}. In the modified version of this model, their masses were estimated as 1840 and 1861 MeV~\cite{Pang:2017dlw}. In the relativistic quark model~\cite{Faustov:2009dlw}, the authors predicted the $2P$ states with masses of  1757 and  1893 MeV. It can be seen that the mass predictions for these radial excited states are still controversial, and only rough range can obtained. According to the theoretical and experimental status of $2P$ states, we tentatively assign the $K_1(1793)$ and $K_1(1861)$ as the $K_1(2P)$ state, and $K_1(1911)$ as the $K_1(2P^\prime)$ state. These assignments also agree with that of the LHCb Collaboration, while the $K_1(1968)$ with a higher mass are not considered because its signal significance is lower than $2\sigma$. The $K_1(1650)$ is also not discussed here due to its lower mass compared with the theoretical predictions.

For the higher excited states, the quark models give significantly different mass spectrum for mesons, which brings difficulties to the clarification for excited kaons. Several reasons and modifications are also discussed in the literature, such as including the screening effects, introducing the unquenched mechanism, or refitting the model parameters. An alternate way is to resort to the Regge trajectories, which indicates that the light mesons could be grouped into some straight lines with similar Regge slopes. In previous works\cite{Tang:Regge,Anisovich:2000kxa,Afonin:2007mesons,Wang:2022pxm}, the Regge trajectories of nonstrange mesons and strangeonium are usually studied, while the studies of excited kaons are relatively few. The recent work predicted the masses of $K_1(3P)$ and $K_1(3P^\prime)$ states are 2126 and  2280  MeV, respectively~\cite{Oudichhya:2023lva}. Considering the experimental status, we tentatively assign the $X(2085)$ resonance as the $K_1(3P)$ state. 

It should be pointed out that these possible assignments for excited kaons are pretty rough. The reason is that the theoretical estimations and experimental data of the masses for these states have large uncertainties and errors, which leads to the obscure interpretations. Moreover, we also need to study the strong decay behaviors to further confirm or exclude the temporary classifications.

	\section{Strong Decays}
	\label{sec:Decay}
 
\subsection{Quark pair creation model}
\label{sec:model}	
\begin{figure}[htpb]
\includegraphics[scale=0.15]{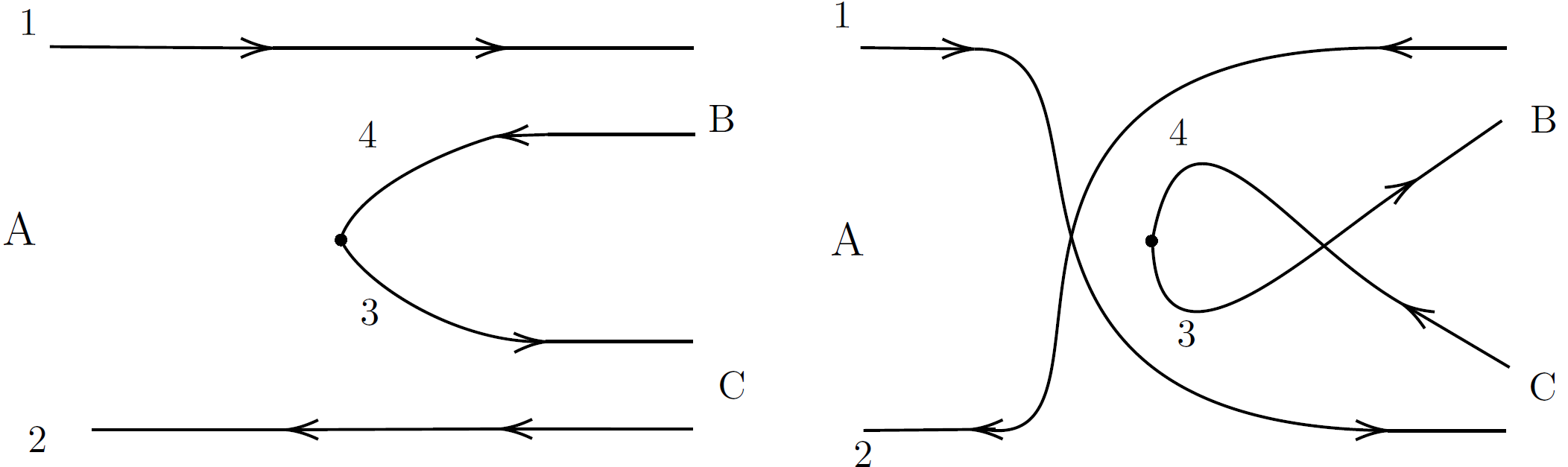}
\vspace{0.0cm}\caption{Two-body-decay diagrams of $A\to BC$ according to the $\tpz$ model, where a pair of quark-antiquark are created to form the final two mesons.
			In the left diagram, Meson $B$ is formed by quark in Meson $A$ combined with the created antiquark, and Meson $C$ is formed by antiquark in Meson $A$ combined with the created quark.
			In the right diagram, Meson $B$ is formed by antiquark in Meson $A$ combined with the created quark, and Meson $C$ is formed by quark in Meson $A$ combined with the created antiquark.
		}\label{fig:feynman}
	\end{figure}

The quark pair creation model, also denoted as the $\tpz$ model, was originally introduced by Micu~\cite{Micu:1968mk} and further developed by Le Yaouanc $et$ $al.$~\cite{LeYaouanc:1972vsx,LeYaouanc:1973ldf,LeYaouanc:alo}, which has been widely used to study the Okubo-Zweig-Iizuka (OZI)-allowed strong decays of various hadrons with considerable success ~\cite{Roberts:1992js,Blundell:1996as,Barnes:1996ff,Barnes:2002mu,
		Close:2005se,Barnes:2005pb,Zhang:2006yj,
		Li:2009rka,Li:2010vx,Lu:2014zua,Pan:2016bac,Lu:2016bbk, Wang:2017pxm,Li:2021qgz,Hao:2019fjg,Xue:2018jvi,Feng:2021igh}.
In this model, the strong decay for a meson takes place by producing a quark-antiquark pair with vacuum quantum number $J^{PC}=0^{++}$. The newly created quark-antiquark pair together with the $q\bar{q}$ in the initial meson, regroups into the two outgoing mesons. The two possible quark rearrangement ways are shown in Fig.~\ref{fig:feynman}. 
	
	Following the conventions in Refs.~\cite{Roberts:1992js,Blundell:1996as}, the transition operator $T$ of the decay  $A\rightarrow BC$ in the $\tpz$ model is given by
	\begin{eqnarray}
		T&=&-3\gamma\sum_m\left< 1m1-m|00\right>\int
		d^3\pbs_3d^3\pbs_4\delta^3(\pbs_3+\pbs_4)\nonumber\\
		&&{\cal{Y}}^m_1\left(\frac{\pbs_3-\pbs_4}{2}\right
		)\chi^{34}_{1-m}\phi^{34}_0\omega^{34}_0b^\dagger_3(\pbs_3)d^\dagger_4(\pbs_4),
	\end{eqnarray}
where the $\gamma$ is a dimensionless parameter corresponding to the production strength of the quark-antiquark pair $q_3\bar{q}_4$ with quantum number $J^{PC}=0^{++}$. The $\pbs_3$ and  $\pbs_4$ are the momenta of the created quark  $q_3$ and  antiquark $\bar{q}_4$, respectively. The $\chi^{34}_{1-m}$, $\phi^{34}_0$, and $\omega^{34}_0$ are the spin, flavor, and color wave functions of $q_3\bar{q}_4$ system, respectively. The solid harmonic polynomial  ${\cal{Y}}^m_1(\pbs)\equiv|\pbs|^1Y^m_1(\theta_p, \phi_p)$ reflects the momentum-space distribution of the $q_3\bar{q_4}$.
	
The $S$ matrix of the process $A\rightarrow BC$ is defined by
	\begin{eqnarray}
		\langle BC|S|A\rangle=I-2\pi i\delta(E_A-E_B-E_C)\langle BC|T|A\rangle,
	\end{eqnarray}
where $|A\rangle$, $|B\rangle$, and $|C\rangle$ are the wave functions of the mock mesons~\cite{Hayne:1981zy}. Then, the transition matrix element $\langle BC|T|A\rangle$ can be written as
	\begin{eqnarray}
		\langle BC|T|A\rangle=\delta^3(\pbs_A-\pbs_B-\pbs_C){\cal{M}}^{M_{J_A}M_{J_B}M_{J_C}}(\pbs),
	\end{eqnarray}
where ${\cal{M}}^{M_{J_A}M_{J_B}M_{J_C}}(\pbs)$ is the helicity amplitude. In the calculation, the partial wave amplitude ${\cal{M}}^{LS}(\pbs)$ is usually adopted, and it relate to the helicity amplitude by a recoupling transformation~\cite{Jacob:1959at},
	\begin{eqnarray}
		{\cal{M}}^{LS}(\pbs)&=&
		\sum_{\renewcommand{\arraystretch}{.5}\begin{array}[t]{l}
				\scriptstyle M_{J_B},M_{J_C},\\\scriptstyle M_S,M_L
		\end{array}}\renewcommand{\arraystretch}{1}\!\!
		\langle LM_LSM_S|J_AM_{J_A}\rangle \nonumber\\
		&&\langle
		J_BM_{J_B}J_CM_{J_C}|SM_S\rangle\nonumber\\
		&&\times\int
		d\Omega\,\mbox{}Y^\ast_{LM_L}{\cal{M}}^{M_{J_A}M_{J_B}M_{J_C}}
		(\pbs). \label{pwave}
	\end{eqnarray}

In the literature, various choices of $\tpz$ models exist, and they typically differ in the quark pair creation vertex, the phase space conventions, and the employed wave functions for initial and final mesons. In present work, we restrict to the simplest vertex as introduced originally by Micu \cite{Micu:1968mk} which assumes a spatially constant pair-production strength $\gamma$, adopt the relativistic phase space, and employ the wave functions  obtained by the relativized quark model~\cite{Godfrey:1985xj}. The flavor wave functions for the mesons are adopted by following the conventions of Refs.~\cite{Barnes:2002mu, Godfrey:1985xj} except for (1) $f_1(1285)=-0.28 n\bar{n}+0.96 s\bar{s}$ and $f_1(1420)=-0.96 n\bar{n}-0.28 s\bar{s}$~\cite{Li:2000dy}, (2) $\eta(1295)=(n\bar{n}-s\bar{s})/\sqrt{2}$ and $\eta(1475)=(n\bar{n}+s\bar{s})/\sqrt{2}$~\cite{Yu:2011ta} with $n\bar{n}=(u\bar{u}+d\bar{d})/\sqrt{2}$. The $\gamma = 0.52$ is obtain by fitting the total width of $\kstarz(1430)$ as the $1^3P_0$ state as our previous work~\cite{Wang:2022pxm}, and the masses of the final mesons are taken from the RPP~\cite{Zyla:2020zbs}. 

Then, the decay width $\Gamma(A\rightarrow BC)$ can be expressed in terms of the partial wave amplitude straightforwardly
	\begin{eqnarray}
		\Gamma(A\rightarrow BC)= \frac{\pi
			|\pbs|}{4M^2_A}\sum_{LS}|{\cal{M}}^{LS}(\pbs)|^2, \label{width1}
	\end{eqnarray}
where $|\pbs|=\sqrt{[M^2_A-(M_B+M_C)^2][M^2_A-(M_B-M_C)^2]}/(2M_A)$, and $M_A$, $M_B$, and $M_C$ are the masses of the mesons $A$, $B$, and $C$, respectively.

\subsection{Results and discussions}
\label{sec:nP}

\begin{figure}[htpb]
\includegraphics[scale=0.22]{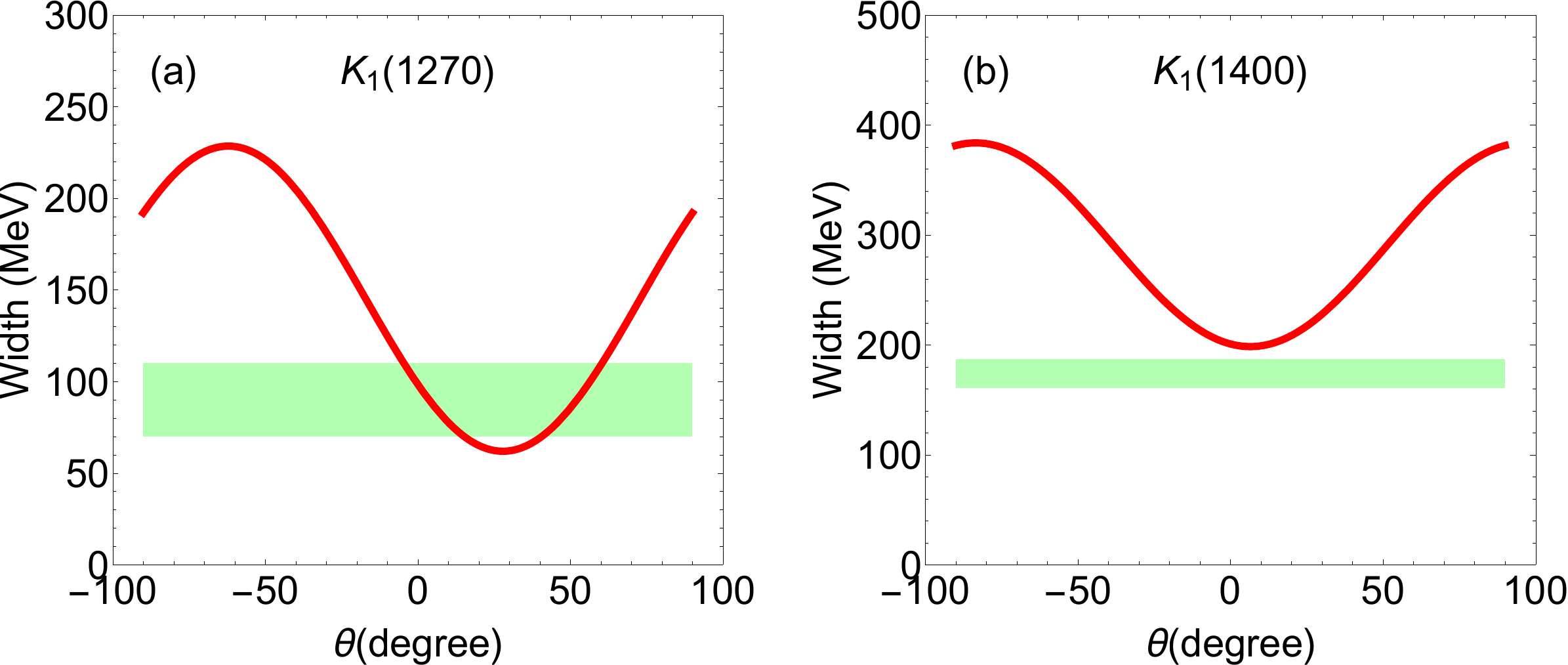}
\vspace{0.0cm}
  \caption{The decay width of 1P varies with the mixing angle $\theta$. On the left is the decay width variation of $K_1(1270)$, and on the right is the decay width variation of $K_1(1400)$.}
  \label{fig:1Ps}
\end{figure}

With above formulas and parameters, one can calculate the strong decays of pure $K(n^1P_1)$ and $K(n^3P_1)$ states. Then, extra mixing angles $\theta_{nP}$ are needed to obtain the decay widths for physical states $K_1(nP)$ and $K_1(nP^\prime)$. For the $K_1(1270)$ and $K_1(1400)$, they are usually regarded as the conventional $K_1(1P)$ and $K_1(1P^\prime)$ states. The $ \theta_{1P}$ dependence of the total decay widths for these two states are shown in Figure~\ref{fig:1Ps}. Compared with the experimental data, we can determine the mixing angle $\theta_{1P}$ locates around $10^{\circ}$. In the literature, the predicted $\theta_{1P}$ has different values such as $\theta_{1P} = $ $25^{\circ}$ or $45^{\circ}$~\cite{Pang:2017dlw}, $\theta_{1P} = $ $45^{\circ}$~\cite{Barnes:2002mu,Blundell:1995au}, $\theta_{1P} = $ $34^{\circ}$ or  $60^{\circ}$~\cite{Cheng:2004kxa,deli:2006kxa,Tayduganov:2012kxa}. It can be seen that the mixture for these low-lying states is quite significant. On the other hand, the mixing angle $\theta_{1P}$ can also be extracted from the potential models, where the spin-orbit interaction provide the mixing mechanism. However, a rather small mixing angle $\theta_{1P}=-4.1^{\circ}$ is obtained in the relativized quark model~\cite{Godfrey:1985xj,Godfrey:1986wj}. This maybe arising from the possible delicate cancellation between the spin-orbit contributions or other dynamic mechanism for the $K(1^1P_1)$ and $K(1^3P_1)$ mixing~\cite{Blundell:1995au,Lipkin:1977uy}. Other processes, such as weak decay of $\tau$ lepton, are encouraged to further investigate the mixing angle $\theta_{1P}$. 

\begin{figure}[htpb]
\includegraphics[scale=0.22]{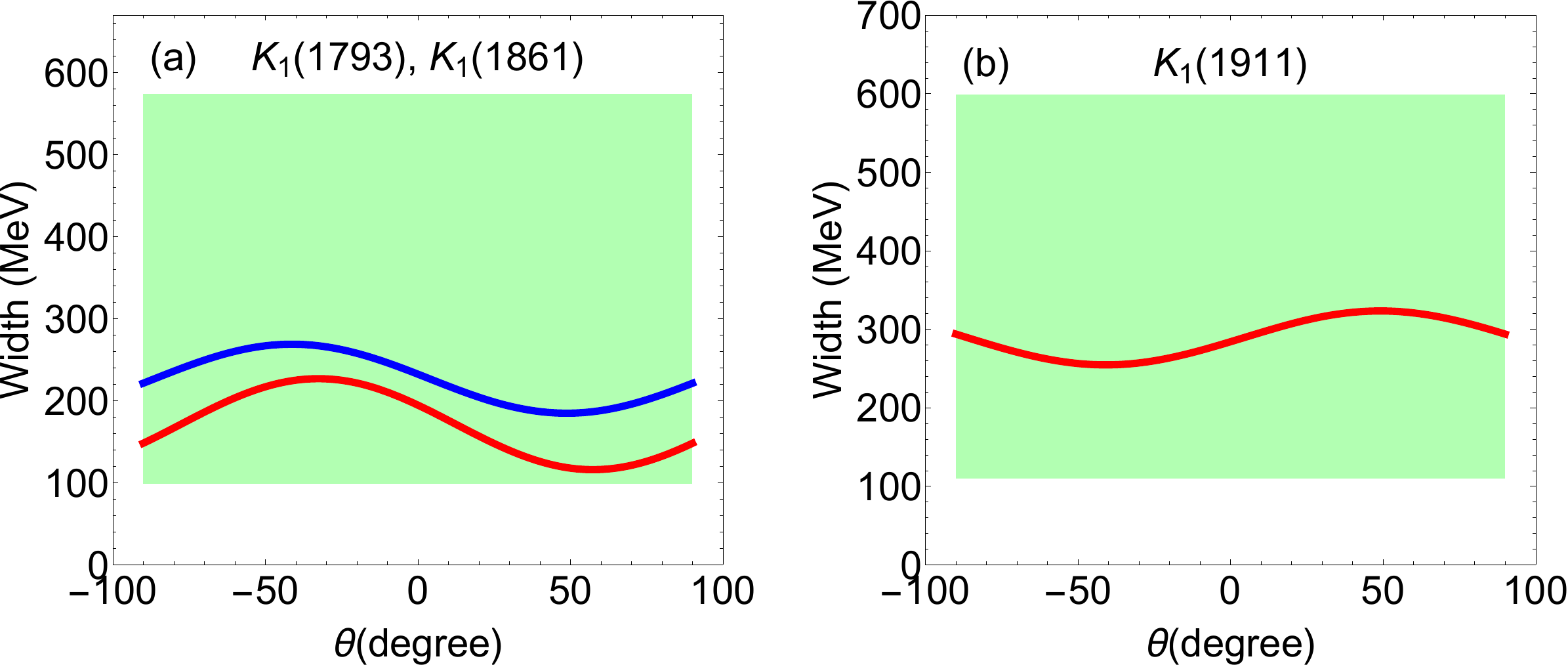}
\vspace{0.0cm}
  \caption{The decay width of 2P varies with the mixing angle $\theta$. On the left is the decay width variation of $K_1(1793)$ (red) and $K_1(1861)$ (blue), and on the right is the decay width variation of $K_1(1911)$.}
  \label{fig:2Ps}
\end{figure}

There are two structures $K_1(1793)$ and $K_1(1861)$ as the candidates of $K_1(2P)$ state, and the structure $K_1(1911)$ may be as a $K_1(2P^\prime)$ state. The predicted mixing angle $\theta_{2P}$ in the relativized quark model is $-14.0^\circ$. With this mixing angle, the strong decay behaviors are calculated and listed in Table~\ref{tab:21P1} $\sim$ \ref{tab:21P1b}. It can be seen that the calculated total decay widths of $K_1(1793)$ and $K_1(1911)$ agree with the experimental data well, while the predicted total width of $K_1(1861)$ is roughly consistent with measurement. Our results suggest that the $K_1(1793)$ and $K_1(1861)$ structures should be the same $K_1(2P)$ state by considering the theoretical and experimental uncertainties. The dominate decay channels for the $K_1(2P)$ state are $\rho K$,  $K^*(1410)\pi$, $\rho K^*$, $\pi K^*$, and $\omega K$, while the $\rho K^*(892)$, $K^*(892)\pi$, $\rho K$, $K^*(1410)\pi$,  and  $\omega K^*(892)$ channels are the main decay modes for $K_1(2P^\prime)$ state. The dependence of total widths of  $K_1(1793)$, $K_1(1861)$, and $K_1(1911)$ on the $\theta_{2P}$ are displayed in Figure~\ref{fig:2Ps}. One can see that when $\theta_{2P}$ varies from $-90^\circ$ to $+90^\circ$, the $K_1(2P)$ and $K_1(2P^\prime)$ assignments can always give  reasonable descriptions of experiments owing to the large experimental uncertainties.  

Both $K_1(2P)$ and $K_1(2P^\prime)$ states are broad resonances with about $200$$\sim$$350$ MeV, which leads to the experimental explorations more challenging. Indeed, the LHCb Collaboration have already carried out two systematic analyses, but the uncertainties of masses and widths  are still so large. Also, the measured channel is $\phi K$, which is relatively smaller than the predicted dominate decay modes. According to present calculations, the $\rho K$, $\rho K^*$, and $\pi K^*$ channels are more suitable to investigate the $K_1(2P)$ and $K_1(2P^\prime)$ states, which can be hunt for  by  the LHCb and BESIII Collaborations in the future.

\begin{table}[h]
	   \begin{center}

	       \caption{Decay widths of $K_1(1793)$ as the $K_1(2P)$ state. The units are in MeV.}
 
		     \label{tab:21P1}
		   \begin{tabular}{c| c c| cc  }
			\hline\hline
			Channel                      & Mode             & $\Gamma_i(2^1P_1)$   & Mode             & $\Gamma_i(2^1P_1)$   \\
\hline
$1^+\rightarrow 1^-+0^- $    & $\rho K$     &  41.13	& $\omega K$   &   13.41    \\
			& $\phi(1020) K$    &   2.25     	& $K^*(892)\pi$    &    27.08       \\
			& $K^*(892)\eta$    &  1.69	   & $K^*(1410)\pi$  &    52.32           \\
     \hline
 $1^+\rightarrow 1^-+1^- $    & $\rho K^*(892)$    & 55.51	& $\omega K^*(892)$    &  18.55\\    
     \hline
$ 1^+\rightarrow0^++0^- $  & $K_0^*(1430)\pi $ & 3.51    &    &   \\
\hline
Total width       &       \multicolumn{4}{c}{215.46}            \\
			\hline
Experiment      &\multicolumn{4}{c}{ 365$\pm$157$^{+138}_{-215}$  \cite{LHCb:2016xxx} } \\
			\hline\hline
			
		\end{tabular}
	\end{center}
\end{table}

\begin{table}[h]
	   \begin{center}

	       \caption{Decay widths of $K_1(1911)$ as the $K_1(2P^\prime)$ state. The units are in MeV.}
 
		     \label{tab:23P1}
		   \begin{tabular}{c| c c| cc  }
			\hline\hline
			Channel                      & Mode             & $\Gamma_i(2P^\prime)$   & Mode             & $\Gamma_i(2P^\prime)$   \\
			\hline
$1^+\rightarrow 1^-+0^- $    & $\rho K$     &  38.32 	& $\omega K$   &     12.47    \\
			& $\phi(1020) K$    &  7.51  	& $K^*(892)\pi$    &    46.29     \\
                & $K^*(892)\eta$    &   0.49 	& $\omega(1420) K$    &       10.20   \\
			 & $K^*(892)\eta'$   & 8.97      & $K^*(1680)\pi$    &       0.51     \\
                & $K^*(1410)\pi$  &      35.94       &    &        \\
  \hline             
$1^+\rightarrow 1^-+1^- $    & $\rho K^*(892)$    &  72.85 & $\omega K^*(892)$    & 23.73 \\ 

\hline
$ 1^+\rightarrow0^++0^- $   & $K_0^*(1430)\pi $    &    8.65 	& $f_0(1370) K$   & 2.36  \\
   \hline
 $ 1^+\rightarrow0^-+2^- $   & $\pi K_2(1770)$  &    $<0.01$  && \\
			\hline
   $1^+\rightarrow 0^-+1^+ $    &  $\eta K_1(1270)$ &  6.73&& \\
\hline
Total width   &       \multicolumn{4}{c}{275.03}            \\
\hline
Experiment      &\multicolumn{4}{c}{276$\pm$50$^{+319}_{-159}$\cite{LHCb:2021xxx} } \\
			\hline\hline
			
		\end{tabular}
	\end{center}
\end{table}	

\begin{table}[h]
	   \begin{center}

	       \caption{Decay widths of $K_1(1861)$ as the $K_1(2P)$ state. The units are in MeV.}
 
		     \label{tab:21P1b}
		   \begin{tabular}{c| c c| cc  }
			\hline\hline
			Channel                      & Mode             & $\Gamma_i(2P)$   & Mode             & $\Gamma_i(2P)$   \\
			\hline
$1^+\rightarrow 1^-+0^- $    & $\rho K$     &  50.39	& $\omega K$   &   16.54   \\
			& $\phi(1020) K$    &  2.80 	& $K^*(892)\pi$    &  35.36    \\
                & $K^*(892)\eta$    &  2.42	 & $K^*(892)\eta'$   & 4.69      \\
                & $K^*(1680)\pi$    &       0.08      & $K^*(1410)\pi$  &     52.00   \\

  \hline             
$1^+\rightarrow 1^-+1^- $    & $\rho K^*(892)$    & 61.08 & $\omega K^*(892)$    & 20.10 \\ 

\hline
$ 1^+\rightarrow0^++0^- $   & $K_0^*(1430)\pi $    &   5.37	& $f_0(1370) K$   &0.16  \\
			\hline
   $1^+\rightarrow 0^-+1^+ $    &  $\eta K_1(1270)$ & 0.16   && \\
\hline
Total width    &       \multicolumn{4}{c}{251.15}            \\
			\hline
			Experiment                  &\multicolumn{4}{c}{149$\pm$41$^{+231}_{-23}$ \cite{LHCb:2021xxx} } \\
			\hline\hline
			
		\end{tabular}
	\end{center}
\end{table}	

\begin{figure}[htpb]
\includegraphics[scale=0.22]{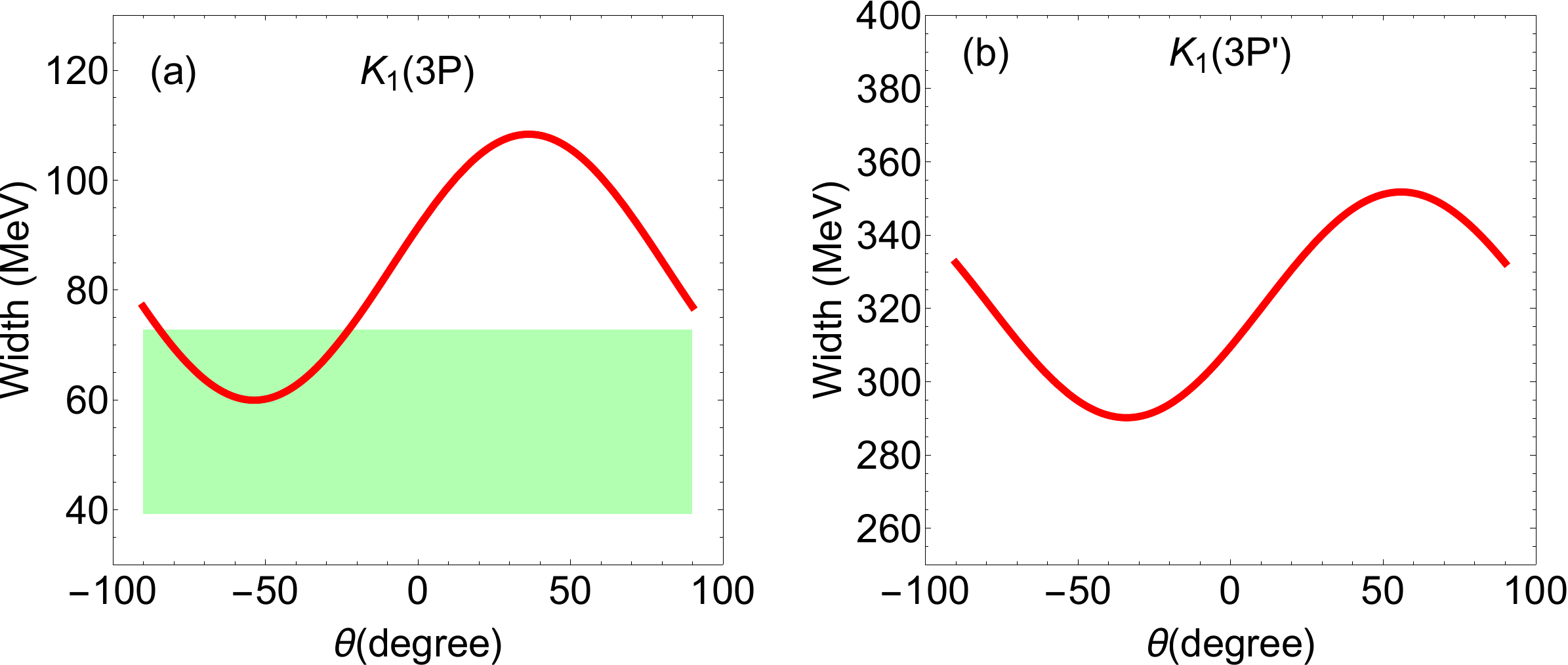}
\vspace{0.0cm}
  \caption{The decay width of 3P varies with the mixing angle $\theta$. On the left is the decay width variation of $X(2085)$, and on the right is the decay width variation of $K_1(2280)$.}
  \label{fig:3Ps}
\end{figure}

As discussed in Sec.~\ref{sec:Mass}, the $X(2085)$ is a good candidate of $K_1(3P)$ state. The strong decay behaviors of $X(2085)$ as the $K_1(3P)$ state are listed in Table~\ref{tab:208531P1}, where the predicted mixing angle $\theta_{3P}=-12.8^\circ$ in the relativized quark model is adopted. The calculated total decay width is about 81 MeV, which is roughly consistent with the measured width 56$\pm$5$\pm$16~MeV by the BESIII Collaboration. Thus, our present results support the assignment of $X(2085)$ as the $K_1(3P)$ state. The dominant decay modes of $K_1(3^1P_1)$ state include $K^*(1410)\pi$, $\rho K^*(892)$, $\omega(1420) K$, $\rho(1450) K$, and $\rho K$ final states, which is helpful for precise measurement  of the $X(2085)$ structure. The dependence on the mixing angle $\theta_{3P}$ for total decay widths is shown in Figure.~\ref{fig:3Ps}. It can be seen that the $K_1(3P)$ state is relatively narrow even the mixing angle varies in a wide range, which is more easily studied than the $K_1(2P)$ and $K_1(2P^\prime)$ states both experimentally and theoretically.   

With the predicted mass 2280 MeV for the $K_1(3P^\prime)$ state, the strong decays are presented in Figure.~\ref{fig:3Ps} versus the mixing angle. Also, the total decay width is predicted to be around 300  MeV, and the mainly decay channels are  $\rho K(1460)$, $\omega(1420) K$, $K^*(892)\pi$, $\rho K_1(1400)$, $K^*(892)a_1(1260)$ and $K^*(892)b_1(1235)$. As a partner of $X(2085)$, we suggest that the BESIII and LHCb Collaborations can search for this state in the final state in future. 

	\begin{table}[h]
	   \begin{center}

	       \caption{Decay widths of $X(2085)$ as the  $K_1(3P)$ state. The units are in MeV.}
 
		     \label{tab:208531P1}
		   \begin{tabular}{c| c c| cc  }
			\hline\hline
Channel         & Mode  & $\Gamma_i(3P)$   & Mode      & $\Gamma_i(3P)$   \\
			\hline
$1^+\rightarrow 1^-+0^- $    & $\rho K$     &    5.07    	& $\omega K$   &        1.60           \\
			& $\phi(1020) K$    &    0.84       	& $K^*(892)\pi$    &     4.26       \\
			& $K^*(892)\eta$    &    0.09  	& $\omega(1420) K$    &       10.90   \\
			& $\rho(1450) K$  &  9.39  & $K^*(892)\eta'$   & 1.78             \\
   
               & $K^*(1680)\pi$    &        3.03       & $K^*(1410)\pi$  &      14.27            \\
               & $K^*(1410)\eta $  &         0.39       &    &                   \\
                  \hline		       
   $ 1^+\rightarrow1^-+1^+ $     & $K^*(892) h_1(1170)$  &  0.26  &$\rho K_1(1270)$ & 5.07    \\
\hline
$ 1^+\rightarrow0^++0^- $     & $a_0(1450) K$    & 1.24     	& $f_0(1370) K$   &   1.01  \\
			& $K_0^*(1430)\pi $    &     0.09    	&   $K_0^*(1430)\eta $  &       0.79       \\
			\hline
   $1^+\rightarrow 0^-+1^+ $    &   $\eta K_1(1400)$ &   $<0.01$ &   $\eta K_1(1270)$ &       0.01 \\
			\hline
    $1^+\rightarrow 1^-+1^- $    & $\phi K^*(892)$   &     0.69	& $\omega K^*(892)$    & 4.13 \\ 
                                    & $\rho K^*(892)$    &   12.24               	 & &         \\
   \hline
 $ 1^+\rightarrow0^-+2^- $     & $\pi K_2(1820) $  &     $<0.01$& $\pi K_2(1770)$  &       0.03    \\
\hline
 $ 1^+\rightarrow0^-+3^- $     & $\pi$ $K_3^*(1780)$ &  3.44  &           &                 \\
\hline
Total width      &       \multicolumn{4}{c}{80.61}            \\
			\hline
			Experiment                  &\multicolumn{4}{c}{56$\pm$5$\pm$16\cite{BESIII:2023xxx} } \\
			\hline\hline
			
		\end{tabular}
	\end{center}
\end{table}

\begin{table}[h]
\begin{center}
\caption{Decay widths of  $K_1(3P^\prime)$ state with initial mass of $2280$~MeV. The units are in MeV.}
 \label{tab:33P11}
 \begin{tabular}{c| c c| cc  }
			\hline\hline
Channel        & Mode             & $\Gamma_i(3P^\prime)$   & Mode   & $\Gamma_i(3P^\prime)$   \\
			\hline
$1^+\rightarrow 1^-+0^- $    & $\rho K$     &  11.43  & $\omega K$   &      3.68     \\
			& $\phi(1020) K$    &   1.91    	& $K^*(892)\pi$    &   22.94      \\
			& $K^*(892)\eta$    &    $<0.01$  	& $\omega(1420) K$    &    25.02  \\
               &  $\phi(1680) K$     &  0.90   &  $\omega(1650) K$     & 8.45  \\
                 &$\rho(1700) K$    & 6.55     &   $\rho(1450) K$    &  17.50   \\
               & $\rho K(1460)$    &   26.78   & $\omega K(1460)$    &  8.31 \\
                 & $K^*(892)\eta'$   & 2.20    & $K^*(892)\pi(1300)$   &  12.75 \\
                 & $K^*(892)\eta(1295)$   & 0.95 & $K^*(1680)\pi$    &   3.92   \\
                 & $K^*(1410)\pi$  &    14.14      & $K^*(1410)\eta$  &    0.12 \\
                 & $K^*(1680)\eta$    &    $<0.01$ &    &                   \\
                 \hline
$1^+\rightarrow 1^-+1^- $    & $\phi K^*(892)$   &  2.10	& $\omega K^*(892)$    & 3.26 \\ 
                       & $\rho K^*(892)$    & 9.94  &      &         \\                
			\hline
$ 1^+\rightarrow0^++0^- $     & $a_0(1450) K$       & 0.70   	& $f_0(1710) K$   &  0.02 \\
                     	& $f_0(1370) K$   &   0.01  & $K_0^*(1430)\pi $    & 0.60 \\
                      &   $K_0^*(1430)\eta $  &  0.29 &   &  \\
                  \hline
 $ 1^+\rightarrow0^-+3^-$ & $\pi$ $K_3^*(1780)$ & 13.93 & $K \rho_3(1690)$  &  6.72    \\
             & $K \omega_3(1670)$ & 8.68  &  &  \\
			\hline 
 $ 1^+\rightarrow0^-+2^- $     & $\pi K_2(1820) $  &  0.77    & $\pi K_2(1770)$  &  0.24  \\
 \hline
   $ 1^+\rightarrow1^-+1^+ $    &$K^*(892)a_1(1260)$   &19.44  &$K^*(892)b_1(1235)$   &  18.02  \\  
                   & $K^*(892) h_1(1170)$  & 0.78  &$K^*(892)f_1(1285)$   & 8.14  \\
                   
                 &$\rho K_1(1270)$ & 15.75  & $\rho K_1(1400)$  &  21.12 \\
			\hline
 $1^+\rightarrow 0^-+1^+ $    &   $\eta K_1(1400)$ & 0.07 &   $\eta K_1(1270)$ & 0.11 \\
\hline
Total width      &       \multicolumn{4}{c}{298.24}            \\
			\hline\hline
		\end{tabular}
	\end{center}
\end{table}

\section{SUMMARY}
\label{sec:summary}

In this work, we study the strong decay behaviors of excited axialvector strange mesons within the quark pair creation model. Our results indicate that the $K_1(1793)/K_1(1861)$ can be regarded as the same $K_1(2P)$ state, and the $K_1(1911)$ is assigned as the $K_1(2P^\prime)$ state. Given the mass, spin-parity, and decay behaviors, the newly observed $X(2085)$ is interpreted as the radially excited $K_1(3P)$ state. Future searches for $X(2085)$ in the $\rho(1450) K$, $\omega(1420)K$, $\pi K^*(1410)$, $\rho K_1(1270)$, and $\rho K^*(892)$ final states can test our present assignment. Moreover, the width of $K_1(3P^\prime)$ state is predicted to be about 300  MeV, the main decay channels of it are $\rho K(1460)$, $\omega(1420) K$, $K^*(892)\pi$, $\rho K_1(1400)$, $K^*(892)a_1(1260)$ and $K^*(892)b_1(1235)$, which can be searched for by future experiments.

For the axialvector strange mesons, we are still unable to study their properties precisely, where significant uncertainties including experimental data, predicted masses, and mixing angles exist. Compared with the broad $K_1(2P)$ and $K_1(2P^\prime)$ states, the narrow $X(2085)$ actually provides an excellent opportunity for us to carefully investigate the axialvector strange mesons. Beside the meson-meson decays, other channels such as $p\bar \Lambda$ and radiative transition may be helpful to draw a final conclusion for the  structure $X(2085)$. More experimental and theoretical efforts on $X(2085)$ are encouraged to better understand its internal structure.   

\subsection*{ACKNOWLEDGMENTS}

We would like to thank Neng-Chang Wei for the fruitful discussions. This work is supported by the National Natural Science Foundation of China under Grant No. 12205075. Q.-F. L\"u is supported by the Key Project of Hunan Provincial Education Department under Grant No. 21A0039 and the Natural Science Foundation of Hunan Province under Grant No. 2023JJ40421.

\end{document}